\documentclass[10.75pt, a4paper]{article}

\usepackage{titlesec}
\titleformat{\section}
  {\bf\sffamily}
  {\thesection. }
  {5pt}
  {\MakeUppercase}
\renewcommand{\thesection}{\Roman{section}} 

\titleformat{\subsection}
  {\bf\sffamily}
  {\thesubsection. }
  {5pt}{}
  
\renewcommand{\thesubsection}{\Alph{subsection}}

\usepackage[affil-it]{authblk} 
\usepackage{etoolbox}
\usepackage{lmodern}

\usepackage{eurosym}

\usepackage[utf8]{inputenc}
\usepackage{geometry}
\usepackage[hidelinks]{hyperref}
\usepackage[citestyle=numeric,style=phys,biblabel=brackets,backend=bibtex,sorting=none,url=false,doi=false, natbib]{biblatex}
\bibliography{bibliography}
\DefineBibliographyStrings{english}{andothers={\itshape et\addabbrvspace al\adddot}}

\usepackage{csquotes}
\usepackage[]{authblk} 
\usepackage{etoolbox}
\usepackage{lmodern}

\usepackage{amsmath}
\usepackage{enumerate}
\usepackage{enumitem}
\usepackage{graphicx}
\usepackage{siunitx}
\usepackage{float}
\usepackage[]{parskip} 

\makeatletter
\patchcmd{\@maketitle}{\LARGE \@title}{\fontsize{16}{19.2}\selectfont\@title}{}{}
\makeatother

\usepackage[normalem]{ulem}

\usepackage{graphicx}
\usepackage{dcolumn}
\usepackage{bm}
\usepackage{tikz}
\usetikzlibrary{math}
\usetikzlibrary{shapes.geometric, arrows}
\usetikzlibrary{positioning}
\usetikzlibrary{shapes,arrows}
\usetikzlibrary{intersections}
\usepackage{csvsimple}
\usepackage{changepage}
\usepackage[tbtags]{mathtools}
\usepackage{siunitx}

\usepackage[normalem]{ulem}
\usepackage[textwidth=\dimexpr\textwidth-2cm\relax]{todonotes}
\makeatletter
\@mparswitchfalse%
\makeatother
\normalmarginpar 

\usepackage{float}
\usepackage[
	nonumberlist, 				
	acronym,      				
	nomain,						
	nopostdot,					
]{glossaries}
\usepackage{glossary-superragged}
\usepackage[hidelinks]{hyperref}
\usepackage{color, colortbl}

\usepackage{wrapfig}

\usepackage{pgfplots}
\usepackage{tikz}
\usetikzlibrary{arrows}
\usepackage{amsmath}
\pgfplotsset{compat=newest}
\usepgfplotslibrary{fillbetween}
\usetikzlibrary{calc}
\def\centerarc[#1](#2)(#3:#4:#5)
{ \draw[#1] ($(#2)+({#5*cos(#3)},{#5*sin(#3)})$) arc (#3:#4:#5); }

\usepackage{amssymb}

\usepackage{nicefrac}

\usepackage{abstract}

\usepackage{multirow}
\usepackage{tabularx}
\usepackage{booktabs}
\usepackage{array}
\usepackage{longtable}
\newcolumntype{L}[1]{>{\raggedright\let\newline\\\arraybackslash\hspace{0pt}}m{#1}}
\newcolumntype{C}[1]{>{\centering\let\newline\\\arraybackslash\hspace{0pt}}m{#1}}
\newcolumntype{R}[1]{>{\raggedleft\let\newline\\\arraybackslash\hspace{0pt}}m{#1}}

\newacronym{3d}{3D}{three dimensional}
\newacronym{am}{AM}{additive manufacturing}
\newacronym{fdm}{FDM}{fused deposition modeling}
\newacronym{ism}{ISM}{in-space manufacturing}
\newacronym{iss}{ISS}{International Space Station}
\newacronym{fcb}{FCB}{Functional Cargo Block}
\newacronym{dem}{DEM}{discrete element method}
\newacronym{md}{MD}{molecular dynamics}
\newacronym{dc}{DC}{direct-current}
\newacronym[plural=PFCs,firstplural=parabolic flight campaigns (PFCs)]{pfc}{PFC}{Parabolic Flight Campaign}
\newacronym{fft}{FFT}{Fast Fourrier Transform}
\newacronym{cad}{CAD}{Computer Assisted Design}
\newacronym{ptfe}{PTFE}{polytetrafluoroethylene}
\newacronym{ps}{PS}{polystyrene}
\newacronym{nasa}{NASA}{National Aeronautics and Space Administration}
\newacronym{esamm}{ESAMM}{Extended Structure Additive Manufacturing Machine}
\newacronym{amf}{AMF}{Additive Manufacturing Facility}
\newacronym{us}{US}{United States}
\newacronym{usa}{USA}{United States of America}
\newacronym{bmgs}{BMGs}{Bulk Metallic Glasses}
\newacronym{esa}{ESA}{European Space Agency}
\newacronym{si}{SI}{International System of Units, abbreviated from French \textit{Syst\`{e}me International (d'unit\'{e}s)}}
\newacronym{dlr}{DLR}{German Aerospace Center}
\newacronym{liggghts}{LIGGGHTS}{\acrshort{lammps} Improved for General Granular and Granular Heat Transfer Simulations}
\newacronym{lammps}{LAMMPS}{Large-scale Atomic/Molecular Massively Parallel Simulator}
\newacronym{sjkr}{SJKR}{Simplified Johnson-Kendall-Roberts}
\newacronym{ded}{DED}{Directed Energy Deposition}
\newacronym{slm}{SLM}{Selective Laser Melting}
\newacronym{sls}{SLS}{Selective Laser Sintering}
\newacronym{eva}{EVA}{Extra-Vehicular Activity}
\newacronym{sem}{SEM}{Scanning Electron Microscopy}
\newacronym{RPM}{RPM}{Ramdom Positioning Machine}
\newacronym{rpm}{rpm}{revolutions per minute}
\newacronym{rise}{RISE}{Research Internships in Science and Engineering}
\newacronym{daad}{DAAD}{German Academic Exchange Service, abbreviated from German \textit{Deutscher Akademischer Austauschdienst}}
\newacronym{fsm}{FSM}{finite-state machine}
\newacronym{ir}{IR}{infrared}
\newacronym{pcbs}{PCBs}{Printed Circuit Boards}
\newacronym{pcb}{PCB}{Printed Circuit Board}
\newacronym{mcr}{MCR}{Modular Compact Rheometer}
\newacronym{sff}{SFF}{Solid Freeform Fabrication}
\newacronym{uv}{UV}{ultraviolet}
\newacronym{abs}{ABS}{acrylonitrile butadiene styrene}
\newacronym{hpde}{HPDE}{high density polyethylene}
\newacronym{pei}{PEI}{polyetherimide}
\newacronym{bff}{BFF}{BioFabrication Facility}
\newacronym{lens}{LENS}{Laser Engineered Net Shaping}
\newacronym{cnc}{CNC}{Computer Numerical Control}
\newacronym{ebf3}{EBF$^3$}{Electron Beam Free-Form Fabrication}
\newacronym{leo}{LEO}{Low Earth Orbit}
\newacronym{pc}{PC}{polycarbonate}
\newacronym{crissp}{CRISSP}{Customisable Recyclable International Space Station Packaging}
\newacronym{Athena}{Athena}{Advanced Telescope for High-ENergy Astrophysics}
\newacronym{lbm}{LBM}{Laser Beam Melting}
\newacronym{bam}{BAM}{Federal Institute for Materials Research and Testing, abbreviated from German \textit{Bundesanstalt f\"{u}r Materialforschung und-pr\"{u}fung}}
\newacronym{pbf}{PBF}{powder bed fusion}
\newacronym{eb}{EB}{Electron Beam}
\newacronym{2d}{2D}{two dimensional}
\newacronym{4d}{4D}{four dimensional}
\newacronym{ft4}{FT4}{Freeman Technology 4 Powder Rheometer}
\newacronym{dsc}{DSC}{Differential Scanning Calorimetry}
\newacronym{pmma}{PMMA}{polymethylmethacrylate}
\newacronym{1g}{$1g$}{gravity on-ground}
\newacronym{mug}{$\mu g$}{microgravity}
\newacronym{bcm}{BCM}{Box Counting Method}
\newacronym{mct}{MCT}{Mode Coupling Theory}
\newacronym{gmct}{gMCT}{granular Mode Coupling Theory}
\newacronym{itt}{ITT}{Integration Through Transients}
\newacronym{mfc}{MFC}{Mass Flow Controller}
\newacronym{ct}{CT}{computed tomography}
\newacronym{xct}{XCT}{X-ray computed tomography}
\newacronym{cv}{CV}{curriculum vitae}
\newacronym{pi}{PI}{principal investigator}
\newacronym{osp}{OSP}{orthogonal superimposed perturbation}
\newacronym{npi}{NPI}{Network Partnering Initiative}
\newacronym{ecsat}{ECSAT}{European Centre for Space Applications and Telecommunications}
\newacronym{eac}{EAC}{European Astronaut Centre}
\newacronym{estec}{ESTEC}{European Space Research and Technology Centre}
\newacronym{fps}{fps}{frames per second}
\newacronym{pdf}{pdf}{probability density function}
\newacronym{al}{Al}{aluminium}
\newacronym{ss}{\textit{SS}}{\textit{Smooth Surface}}
\newacronym{rs}{\textit{RS}}{\textit{Rough Surface}}
\newacronym{rcp}{rcp}{random close packing}
\newacronym{iop}{IoP UvA}{Institute of Physics of the University of Amsterdam}
\newacronym{mp}{MP}{Institute of Material Physics for Space}
\newacronym{elgra}{ELGRA}{European Low Gravity Research Association}
\newacronym{zarm}{ZARM}{Center of Applied Space Technology and Microgravity}
\newacronym{piv}{PIV}{particle image velocimetry}

\usepackage[framemethod=tikz]{mdframed}
\usepackage{lipsum}
\usepackage{enumitem}
\usepackage{mdframed}
\usepackage{amsmath}
\usepackage{physics}
\usepackage{setspace}
\usepackage{mathtools}
\usepackage{hyperref}
\usepackage{graphicx}
\usepackage{caption}
\usepackage{subcaption}
\usepackage{wasysym}
\usepackage{bm}
\usepackage{xcolor}
\usepackage[english]{babel}
\usepackage[nottoc]{tocbibind}
\usepackage[dvipsnames]{xcolor}
\usepackage{gensymb}

\usepackage[most]{tcolorbox}
\newtcolorbox{myBox}[3][]{
arc=2mm,
lower separated=true,
fonttitle=\bfseries,
colbacktitle=gray!10,
coltitle=black!50!black,
enhanced,
colframe=gray!10,
colback=gray!10,
title=#2,#1}

\usepackage{dirtytalk}
\usepackage{tcolorbox}
\usepackage[font=footnotesize]{caption}
\newtcolorbox{mybox}[1]{colback=green!6!white,colframe=black!75!black,fonttitle=\bfseries,title=#1}
\newtcolorbox{mybox2}{colback=red!5!white,colframe=red!75!black}

\usepackage{pifont}

\usepackage{soul,xcolor}
\setstcolor{red}

\usepackage{xcolor,hyperref}
\hypersetup{
   colorlinks,
   linkcolor={blue!50!black},
   citecolor={blue!50!black},
   urlcolor={blue!80!black}
} 

\definecolor{mycolor}{rgb}{0.122, 0.435, 0.698}

\usepackage{textgreek}

\title{Rheology of Two-Dimensional Dilute Emulsions}

\author[1]{Thomas Appleford 
\footnote{t.appleford@uva.nl}
}

\affil[1]{Van der Waals-Zeeman Institute, Institute of Physics, University of Amsterdam, The Netherlands}

\author[2]{Vatsal Sanjay
\footnote{ORCID: 0000-0002-4293-6099}
}

\affil[2]{CoMPhy Lab, Department of Physics, Durham University, United Kingdom
}

\author[1]{Maziyar Jalaal 
\footnote{m.jalaal@uva.nl, ORCID: 0000-0002-5654-8505}
}

\begin{document}
    \definecolor{brickred}{rgb}{0.8, 0.25, 0.33}
    \definecolor{darkorange}{rgb}{1.0, 0.55, 0.0}
    \definecolor{persiangreen}{rgb}{0.0, 0.65, 0.58}
    \definecolor{persianindigo}{rgb}{0.2, 0.07, 0.48}
    \definecolor{cadet}{rgb}{0.33, 0.41, 0.47}
    \definecolor{turquoisegreen}{rgb}{0.63, 0.84, 0.71}
    \definecolor{sandybrown}{rgb}{0.96, 0.64, 0.38}
    \definecolor{blueblue}{rgb}{0.0, 0.2, 0.6}
    \definecolor{ballblue}{rgb}{0.13, 0.67, 0.8}
    \definecolor{greengreen}{rgb}{0.0, 0.5, 0.0}
    \begingroup
    \sffamily
    \date{}
    \maketitle
    \endgroup
    
    \begin{abstract}
        The single droplet under shear is a foundational problem in fluid mechanics.
        In computational fluid dynamics (CFD), the two-dimensional (2D) problem offers advantages in both computational efficiency and relevance, 
        yet it lacks a comprehensive theoretical treatment equivalent to that of the three-dimensional (3D) case.
        In this brief note, we present an analytical treatment of this problem beginning with a derivation of the Lamb solution for 2D Stokes flows, and then applying it to obtain the flow field around a droplet in a purely extensional flow.
        Using these flow fields, expressions are obtained for the apparent viscosity, $\mu^*$, of a dilute 2D emulsion as well as a small deformation theory.
        We show that in the limit of both zero Reynolds number ($\text{Re} \to 0$) in which inertia is negligible and zero capillary number ($\text{Ca} \to 0$) in which surface tension prevents any droplet deformations,  
        $\mu^* = \mu( 1 + f(\lambda) \phi) + \mathcal{O}(\phi^2)$
        with
        $f(\lambda) = (2\lambda + 1)/(\lambda + 1)$
        where $\lambda$ is the ratio of the droplet viscosity to the matrix viscosity and $\phi$ is the area fraction covered by the suspended phase.
        In addition, we show that in the capillarity dominated regime ($\text{Ca} \ll 1$), the steady state value, $D_T^\infty$, of the Taylor deformation parameter obeys $D_T^\infty = g(\lambda)\,\text{Ca}$, where Ca is the capillary number and $g(\lambda) = 1$.
        This contrasts with the 3D case, where $g(\lambda)$ depends on $\lambda$.
        These results are then validated through direct numerical simulations (DNS) of a single droplet at the centre of a periodic square box undergoing simple shear.
        The DNS and theory are compared across a wide range of viscosity ratios
        ($0.01 \leq \lambda \leq 100$)
        and area fractions
        ($0.002 \leq \phi \leq 0.125$)
        and they are found to agree asymptotically in the limit $\phi \to 0$ for $\text{Ca} \ll 1/(\lambda + 1)$.
        Our results provide a basic theoretical framework for interpreting 2D droplet simulations and provide clear benchmarks for CFD codes in 2D.
        \\
         \\
        \textbf{keywords: Droplets $|$ Emulsions $|$ Surface Tension}
    \end{abstract}
    
    \section{Introduction}
        \label{sec:into}
        An emulsion is a heterogeneous mixture of two immiscible fluids, where one phase (the dispersed phase) consists of droplets dispersed in the other phase (the continuous phase), often stabilized by a surfactant.
        Emulsions typically exhibit complex rheological properties, such as elasticity or yield stress, even when neither of the phases independently demonstrates these characteristics. 
        These additional properties make emulsions crucial in a variety of scientific and industrial contexts~\cite{derkach2009rheology}, including food science, cosmetics, pharmaceuticals, oil recovery~\cite{rallison1984deformation}, and the mixing of viscous liquids~\cite{tjahjadi1991stretching}.
        The rheological behaviour of emulsions, particularly in the dilute limit, depends on multiple factors, including the size distribution and volume fraction of the droplets, surface tension, the presence of surfactants, and the relative viscosities of the phases.
        However, perhaps more fundamentally, the properties of an emulsion are a result of interactions between the droplets and the surrounding flow.
        Thus, understanding individual droplets under shear remains central to the study of emulsion physics.
        
        In simple shear, a droplet deforms as a result of the imbalance between viscous forces and surface tension. 
        The droplet elongates in the direction of the shear, with the extent of the deformation primarily influenced by the shear rate, the viscosity of each phase, and the droplet's surface tension. 
        At moderate shear rates, the droplet attains a stable shape; however, at excessively high shear rates, it ruptures.
        The first theoretical treatment of the droplet under shear is found in the seminal works of Taylor~\cite{Taylor1932,Taylor1934}, in which Lamb's general solution of the Stokes equations~\cite{Lamb} is used to derive an analytical solution for purely extensional flow around a non-deformable droplet. 
        Taylor then generalises Einstein's formula~\cite{einstein1905neue} for the viscosity of a dilute suspension of rigid particles to the case of viscous droplets, as well as deriving a small deformation theory.
        
        Subsequent studies have sought to extend these results to larger, time-dependent deformations~\cite{Cox1969, ChaffeyBrenner, Maffettone1998, rallison1984deformation, Rallison_1980}, confinement effects~\cite{Ioannou2016}, surfactant effects~\cite{stone1990effects,vlahovska2009small,soligo2020deformation}, and droplet breakup~\cite{Li2000,bentley1986experimental}.
        More recent computational studies have examined droplets in shear flows where the phases exhibit non-Newtonian rheological properties ~\cite{Guido2011-ug, Li2000_BI, wang2022droplet}.
        The topic has been reviewed in the literature many times, notably in the articles by Acrivos~\cite{acrivos1983breakup}, Rallison~\cite{rallison1984deformation}, and Stone~\cite{stone1994dynamics}. More recent reviews on related topics are available in~\cite{minale2010models, reboucas2025modeling}.
        
        Although most computational studies focus on three-dimensional (3D) systems, the high computational cost has led researchers to pursue more efficient alternatives.
        Among these, simulations in two dimensions (2D) have emerged as a more tractable option, requiring significantly fewer computational resources.
        These have been applied extensively to single droplet systems \cite{yang}, as well as to dilute and dense emulsions ~\cite{foglino2017flow, negro2023yield}.
        However, when working in reduced dimensions, the question inevitably arises as to whether all the essential physics of the 3D system is faithfully represented by its lower-dimensional analogue. 
        It has been argued, for example, that in 2D Stokes flows, droplets do not become unstable and rupture, but instead always reach a stable steady shape \cite{wagner2003role}.
        This is in stark contrast with the 3D case where droplets are known to rupture even in the limit of zero inertia.
        In view of these differences between the 2D and 3D cases, it is unclear to what extent the many classic results of Taylor carry over to the 2D case, either quantitatively or qualitatively.
        Even for a single droplet under shear, there is presently no 2D theoretical treatment that is directly comparable in scope to Taylor's 3D analysis, so this question remains difficult to answer.
        There has been some important progress including the classic works of Richardson~\cite{richardson1968two, richardson1973two} on 2D bubbles, the study of Buckmaster \& Flaherty~\cite{buckmaster1973bursting} on 2D drops, 
        as well as the work of Brady~\cite{BRADY1983} on the apparent viscosity of particular suspensions in arbitrary dimensions.
        Nevertheless, there is as yet no unified framework that yields both the 2D steady-state shape and the apparent viscosity of a dilute emulsion of droplets across a range of viscosity ratios.
        Here we take a step towards such a framework.
        We first derive a 2D analogue of Lamb's solution to the Stokes equations and then use it to solve for purely extensional flow around a viscous droplet, in the regime of strong capillarity where droplet deformations are negligible.
        We then use this solution to obtain 2D analogue expressions for the apparent viscosity of non-deformable droplets, before developing a small-deformation theory, valid over a range of viscosity ratios.
        The 2D results bear strong resemblances to their 3D counterparts, but also highlight some important differences, particularly in regard to the numerical prefactors relating the apparent viscosity to the coverage fraction of the suspended phase, as well as in the dependence of the deformation theory on the viscosity ratio.
        These differences are crucial for researchers developing 2D codes for solving multiphase flows and who require clear and consistent benchmarks for their simulations.
        These benchmarks are then demonstrated with our own direct numerical simulations (DNS).

    \section{Problem Statement}
        \label{sec:prob}
        \subsection{Problem Configuration}
\begin{figure}[h]
                \centering\includegraphics[width=0.45\linewidth]{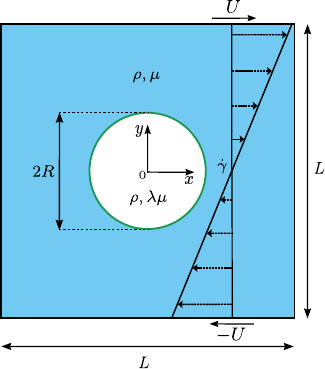}
        
                \caption
                {
                    The configuration of the the simulations of the droplet under shear problem. 
                    A droplet radius $R$ is placed at the centre of a square domain of side length $L$ such that the area fraction is $\phi = \pi R^2/L^2$. 
                    The upper and lower walls move with speeds $\pm \dot{\gamma}L/2$ respectively such that the average strain rate is $\dot{\gamma}$.
                    The domain is left-right periodic.
                    The fluids have equal density, $\rho$, but a viscosity ratio $\lambda$.
                }
                \label{fig:configuration}
            \end{figure}

            A circular droplet of radius $R$ is placed at the origin, at the centre of a square domain of side length $L$ (in our theoretical treatment of the problem in Section \ref{sec:analytics}, we will assume $L/R \to \infty)$.
            The components of the velocity field, 
            $\bm{u} = (u,v)$, 
            are initialised as $u(x, y) = \dot{\gamma}y$, $v(x, y) = 0$, so that the system is immediately in a state of flow (see figure~\ref{fig:configuration}).
            The upper and lower edges of the domain can be thought of as parallel plates moving in opposite directions to maintain the shear flow between them.
            Accordingly, the boundary conditions imposed on the upper and lower edges are 
            $u(y = \pm L/2) = \pm U$
            where 
            $U = \dot{\gamma}L/2$
            such that the average strain rate between the two plates is given by $\dot{\gamma}$.
            On the left and right edges of the domain, we impose periodic boundary conditions such that the system has periodicity $L$.
            The fluids inside and outside the droplet have an equal density $\rho$.
            The viscosity of the matrix is denoted by $\mu$, whilst the viscosity of the fluid in the droplet is $\lambda \mu $. 
            The coefficient of surface tension at the droplet interface is constant and denoted by $\sigma$.
        
        \subsection{Equations of Motion}
            Since the system comprises incompressible fluids, the continuity equation in each phase takes the following form:
            \begin{align}
                \bm{\nabla \cdot u} &= 0.
            \end{align}
            Furthermore, each fluid is Newtonian. Thus, the momentum equation for the system is the incompressible Navier-Stokes equation, which reads:
            \begin{align}
                \rho
                \left(
                \pdv{\bm{u}}{t}
                +
                (\bm{u \cdot \nabla}) \bm{u}
                \right)
                &=
                -
                \bm{\nabla} p
                +
                \bm{\nabla \cdot}\left(2\mu\left(f + [1-f]\lambda\right)\bm{E}\right)
                +
                \sigma\kappa \bm{\hat{n}} \delta_S,
                \label{eqn:navier_stokes}
            \end{align}
            where $f = f(x, y)$ is an indicator function which takes the value $0$ inside the droplet and $1$ outside, $\bm{E} = \left(\bm{\nabla u} + \left(\bm{\nabla u}\right)^T\right)/2$ is the symmetric part of the velocity gradient tensor, $\kappa$ is the curvature of the interface, $\bm{\hat{n}}$ is the unit normal to the interface and $\delta_S$ is a Dirac delta distribution centred on the interface.
            We make the equations of motion dimensionless by re-scaling the time, length, and pressure variables by their characteristic values. 
            These are the shear flow time scale, $1/\dot{\gamma}$, the initial radius, $R$, and the viscous stress scale, $\mu\dot{\gamma}$.
            The dimensionless groups that appear are the Reynolds number, Re, and the capillary number, Ca, which are defined as follows:
            \begin{align}
                \text{Re}
                &=
                \frac{\rho\dot{\gamma} R^2}{\mu},
                \\
                \text{Ca}
                &=
                \frac{\mu \dot{\gamma} R}{\sigma}.
            \end{align}
            The equations then read:
            \begin{align}
                \text{Re}
                \left(
                \pdv{\bm{u}^*}{t^*}
                +
                (\bm{u}^* \bm{\cdot \nabla}^*) \bm{u}^*
                \right)
                &=
                -
                \bm{\nabla}^* p^*
                +
                \bm{\nabla}^* \bm{\cdot}\left(2\left(f + [1-f]\lambda\right)\bm{E}^*\right)
                +
                \frac{1}{\text{Ca}}\kappa^* \bm{\hat{n}} \delta_S^*.
                \label{eqn:navier_stokes_nondim}
            \end{align}   
            where all the starred quantities are dimensionless.
            The Reynolds number quantifies the ratio of inertial and viscous forces within the system, whilst the capillary number quantifies the relative size of the viscous forces and capillary forces.
            Since we imagine the droplet to be small, we focus our attention exclusively on the Stokesian limit with small $\text{Re}$ (more specifically $\text{Re} = 0.01$), where inertial effects are negligible, and the regime of small $\text{Ca}$ (more specifically $\text{Ca} \ll 1$) in which droplets undergo finite deformations and do not rupture.
            In simulations, we typically, unless otherwise stated, fix the ratio $L/R = 20$ such that the interactions between the droplet and the wall, as well as self-interactions with its periodic image are negligible. 
            This fixes the area fraction of the suspended phase $\phi = \pi R^2/L^2 \approx 0.008$, such that when viewed as a periodic droplet emulsion, the system can be considered dilute.
            In other simulations, where $\phi$ is explicitly varied, we vary $L/R$ between 5 and 20 such that $\phi$ varies from $\phi \approx 0.008$ to $\phi \approx 0.125$.

    \section{Analytical Results in the Stokesian Limit}
        \label{sec:analytics}
        \subsection{The Lamb Solution for Two-Dimensional Stokes Flows}
            As previously discussed, we will focus exclusively on the Stokesian limit ($\text{Re} \to 0$).
            In this regime, the full system of equations reduces to the Stokes equations:
            \begin{align}
                \bm{\nabla \cdot u}
                &=
                0
                \\
                \bm{\nabla} p
                &=
                \mu \nabla^2 \bm{u}.
                \label{eqn:stokes}
            \end{align}
            whose general solution is:
            \begin{align}
                \bm{u}
                &=
                \sum_{n=-\infty}^{\infty}
                \left\{
                \bm{\nabla} \phi_n
                + 
                \bm{\nabla} \times (\bm{x}\chi_n)
                +
                \frac{n+2}{4 \mu n(n+1)}
                r^2 \bm{\nabla} p_n
                -
                \frac{1}{2\mu(n+1)}p_n \bm{x}
                \right\}
                \label{eqn:lamb}
                \\
                p 
                &=
                \sum_{n=-\infty}^\infty
                p_n
                \label{eqn:lamb2}
            \end{align}
            where $p_n$, $\phi_{n}$ and $\chi_n$ are arbitrary solid circular harmonics of degree $n$
            (see Appendix~\ref{appendix:analytics} for the derivation details).
            In addition, for the purpose of calculation, we will assume that we are in the limit of infinite surface tension ($\text{Ca} \to 0$), where the droplet remains circular.
            We then divide the ambient space into two separate regions, namely the interior, $\Omega^{\text{int}}$, and exterior, $\Omega^{\text{ext}}$, of the circular droplet, before imposing boundary conditions at the interface.
            Thus, we choose different forms of $p_n$ depending on the region.
            From here on in, we shall use the superscripts $^{\text{int}}$ and $^{\text{ext}}$ to indicate that a variable is defined in $\Omega^{\text{int}}$ and $\Omega^{\text{ext}}$ respectively.
        
        \subsection{Non-Deformable Droplet in a Purely Extensional Flow}
            The undisturbed simple shear flow described in Section \ref{sec:prob} can be written in the form $\bm{u}^\infty = \bm{E}^\infty \bm{\cdot x} + \bm{R}^\infty \bm{\cdot x}$ where $\bm{E}^\infty$ is a symmetric tensor representing the purely extensional part of the undisturbed flow and $\bm{R}^\infty$ represents the solid-body rotational part.
            The rotational part, $\bm{R}^\infty \bm{\cdot x}$, of the undisturbed flow is also the solution to rotational flow around a single circular droplet.
            Thus, by linearity of the Stokes equations, the disturbance field in the case of simple shear will be the same as in the case of purely extensional flow.
            To this end, we chose $p_n$, $\phi_{n}$ and $\chi_n$ to be linear in $\bm{E}^\infty$ and not depend on $\bm{R}^\infty$.
            For the flow on the interior of the droplet, we choose:
            \begin{align}
                p_2 
                &=
                a_2 \lambda \mu 
                \left(\frac{r}{R}\right)^2
                \frac{\bm{x \cdot}  \bm{E}^\infty \bm{\cdot x}}{r^2}
                \label{eqn:ansatz_interior}
                \\
                \phi_2 
                &=
                b_2  
                R^2
                \left(\frac{r}{R}\right)^2
                \frac{\bm{x \cdot}  \bm{E}^\infty \bm{\cdot x}}{r^2},
                \label{eqn:ansatz_interior2}
            \end{align}
            meanwhile, on the exterior, we choose:
            \begin{align}
                p_{-2}
                &=
                a_{-2} \mu 
                \left( \frac{R}{r} \right)^2
                \frac{\bm{x \cdot}  \bm{E}^\infty \bm{\cdot x}}{r^2}
                \label{eqn:ansatz_exterior}
                \\
                \phi_{-2} 
                &=
                b_{-2} R^2
                \left( \frac{R}{r} \right)^2
                \frac{\bm{x \cdot}  \bm{E}^\infty \bm{\cdot x}}{r^2}.
                \label{eqn:ansatz_exterior2}
            \end{align}
            where $a_{-2}$, $b_{-2}$, $a_2$, and $b_2$ are constants determined by the boundary conditions of the problem.
            For all other values of $n$, we take $p_n$, $\phi_{n}$ and $\chi_n$ to be zero.
            We then substitute 
            Eq. \eqref{eqn:ansatz_interior} 
            ,
            Eq. \eqref{eqn:ansatz_interior2}
            ,
            Eq. \eqref{eqn:ansatz_exterior}
            and
            Eq. \eqref{eqn:ansatz_exterior2}
            into 
            Eq. \eqref{eqn:lamb} 
            and
            Eq. \eqref{eqn:lamb2} 
            to obtain general expressions for the velocity and pressure fields, then impose the boundary conditions.
            The boundary conditions are
            the continuity of the tangential stress at the boundary, the continuity of the velocity at the boundary, and the vanishing of the velocity component normal to the boundary, which can be stated as follows:
            \begin{align}
                \left[
                \bm{\hat{n} \cdot}  
                (\bm{\sigma}^\text{int} - \bm{\sigma}^\text{ext})
                \bm{\cdot \epsilon \cdot \hat{n}} 
                \right]_{r = R}
                &=
                \bm{0}
                \\
                \left[
                \bm{u}^\text{int} - \bm{u}^\text{ext}
                \right]_{r = R}
                &= 
                \bm{0}
                \\
                \left[
                \bm{u}^\text{int}\bm{\cdot \hat{n}} 
                \right]_{r = R}
                &= 
                0,
            \end{align}
            where $\bm{\sigma} = -p\,\bm{\delta} + \bm{\tau}$ is the Cauchy stress tensor, $\bm{\delta}$ is the two-dimensional Kronecker symbol, $\bm{\epsilon}$ is the two-dimensional Levi-Civita symbol, and $\bm{\hat{n}}  = \bm{x}/r$ is the unit normal vector.
            It can be shown that the boundary conditions require:
            \begin{align}
                a_2 
                &=
                \frac{6}{\lambda +1} \\
                b_2 
                &=
                -\frac{\lambda + 2}{2(\lambda +1)}\\
                a_{-2} 
                &= 
                -\frac{2(2\lambda+1)}{\lambda +1}\\
                b_{-2} 
                &= 
                - \frac{\lambda}{2(\lambda +1)}
            \end{align}
            The solution to the droplet under shear problem then reads:
            \begin{align}
                p^{\text{int}}
                &=
                \frac{6\lambda \mu}{\lambda + 1} 
                \left(\frac{r}{R}\right)^2
                \frac{\bm{x \cdot}  \bm{E}^\infty \bm{\cdot x}}{r^2}
                \\
                \bm{u}^{\text{int}}
                -
                \bm{u}^\infty
                &=
                -\frac{\lambda + 2}{\lambda +1}
                \bm{E}^\infty \bm{\cdot x}
                +
                \frac{1}{\lambda +1}
                \left( 
                \frac{r}{R}
                \right)^2
                \left( 
                2 \bm{E}^\infty \cdot \bm{x}
                -
                \frac{\bm{x \cdot}  \bm{E}^\infty \bm{\cdot x}}{r^2} \bm{x}
                \right)
                \label{eqn:u_int}
                \\
                p^{\text{ext}}
                &=
                -\frac{2(2\lambda+1)\mu}{\lambda +1}  
                \left( \frac{R}{r} \right)^2
                \frac{\bm{x \cdot}  \bm{E}^\infty \bm{\cdot x}}{r^2}
                \\
                \bm{u}^{\text{ext}}
                -
                \bm{u}^\infty
                &=
                -
                \frac{2\lambda+1}{\lambda +1}
                \left( 
                \frac{R}{r}
                \right)^2
                \frac{\bm{x \cdot}  \bm{E}^\infty \bm{\cdot x}}{r^2} \bm{x}
                -
                \frac{\lambda}{\lambda +1}
                \left( 
                \frac{R}{r}
                \right)^4
                \left( 
                \bm{E}^\infty \cdot \bm{x}
                -
                2 \frac{\bm{x \cdot}  \bm{E}^\infty \bm{\cdot x}}{r^2} \bm{x}
                \right)
                \label{eqn:u_ext}
            \end{align}
            In the limit $\lambda \to \infty$, we find:
            \begin{align}
                p^{\text{int}}
                &=
                6\mu
                \left(\frac{r}{R}\right)^2
                \frac{\bm{x \cdot}  \bm{E}^\infty \bm{\cdot x}}{r^2}
                \\
                \bm{u}^{\text{int}}
                &=
                \bm{0}
                \\
                p^{\text{ext}}
                &=
                -4\mu 
                \left( \frac{R}{r} \right)^2
                \frac{\bm{x \cdot}  \bm{E}^\infty \bm{\cdot x}}{r^2}
                \\
                \bm{u}^{\text{ext}}
                -
                \bm{u}^\infty
                &=
                -
                2
                \left( 
                \frac{R}{r}
                \right)^2
                \frac{\bm{x \cdot}  \bm{E}^\infty \bm{\cdot x}}{r^2}\bm{x}
                -
                \left( 
                \frac{R}{r}
                \right)^4
                \left( 
                \bm{E}^\infty \cdot \bm{x}
                -
                2 \frac{\bm{x \cdot}  \bm{E}^\infty \bm{\cdot x}}{r^2}\bm{x}
                \right)
            \end{align}
            which is precisely the result for purely extensional flow around a cylinder derived in \cite{Brenner_1981}.

        \subsection{The Apparent Viscosity of a Two-Dimensional Emulsion of Viscous Droplets}
            We continue by deriving an analytical expression for the apparent viscosity, $\mu^*$, of a dilute $(\phi \ll 1)$ emulsion comprising non-deformable droplets.
            The apparent viscosity will take the form:
            \begin{align}
                \mu^*
                &=
                \mu
                \left(
                    1 + f(\lambda) \phi
                \right)
                +
                \mathcal{O}(\phi^2)
                \label{eqn:einstein_viscosity}
            \end{align} 
            where, as before, $\phi$ is the  area fraction of the suspended phase, and $f(\lambda)$ is some function of the viscosity ratio, $\lambda$, to be determined here.
            The function $f(\lambda)$ is readily determined by computing the total rate of energy dissipation, $\Phi$, of the droplet under shear system, which can be written as \cite{batchelor2000introduction,batchelor1970stress}:
            \begin{align}
                \Phi
                &=
                \frac{1}{L^2}
                \int 
                \bm{\sigma} \bm{:} \bm{E} 
                \dd{A}
                \\
                &=
                \frac{1}{L^2}
                \int_{\Omega^{\text{int}}}
                2\lambda \mu \bm{E} \bm{:} \bm{E} 
                \dd{A}
                +
                \frac{1}{L^2}
                \int_{\Omega^{\text{ext}}}
                2\mu \bm{E} \bm{:} \bm{E} 
                \dd{A}.
            \end{align}
            By definition of the apparent viscosity, we may also write:
            \begin{align}
                \Phi
                &=
                \mu^* \dot{\gamma}^2
            \end{align}
            where $\Omega = \Omega^{\text{int}} \cup \Omega^{\text{ext}}$ is the entire domain. 
            Thus, we may write:
            \begin{align}
                \frac{\mu^*}{\mu}
                &=
                \frac{2}{\dot{\gamma}^2}
                \left(
                    \int_{\Omega^{\text{int}}}
                    \lambda \bm{E} \bm{:} \bm{E} 
                    \dd{A}
                    +
                    \int_{\Omega^{\text{ext}}}
                    \bm{E} \bm{:} \bm{E} 
                    \dd{A}
                \right)
                \label{eqn:einstein_dissipation}
            \end{align}
            It can be shown using the results in Eq. \eqref{eqn:u_int} and Eq. \eqref{eqn:u_ext} that:
            \begin{align}
                f(\lambda)
                &=
                \frac{2\lambda + 1}{\lambda + 1}.
                \label{eqn:apparent_2D}
            \end{align}
            This is in contrast with the 3D result \cite{Taylor1932}, which reads:
            \begin{align}
                f^{3D}(\lambda)
                &=
                \frac{5\lambda + 2}{2(\lambda + 1)}.
            \end{align}
            It will be noted that in the limit $\lambda = 0$, in which the droplet is inviscid, we have $f(0) = f^{3D}(0) = 1$, that is, the 2D and 3D cases give identical results.
            On the other hand, in the limit $\lambda \to \infty$, when the droplet would behave analogously to a solid particle \cite{galeano2021capillary}, we have $f(\lambda \to \infty) = 2$ but $f^{3D}(\lambda \to \infty) = 5/2$.
            The deviation, therefore, between the 2D and 3D results becomes most apparent when the viscosity ratio, $\lambda$ is large, with the increase in viscosity in the 2D case being smaller than that in the 3D case.
            We also note that in the limit $\lambda \to \infty$, our result is consistent with the result derived in~\cite{BRADY1983}.

        \subsection{Small Deformation Theory}
            We use the analytical solution to study small deformations of the droplet.
            In finding the analytical solution, we imposed the condition that the tangential component of the radial traction vector be continuous at the surface of the droplet.
            The difference in normal component, however, is discontinuous at the surface of the droplet.
            In particular, we find the difference,
            $
                \Delta \bm{\Sigma}_\mu
                :=
                \left[
                \bm{\hat{n} \cdot } 
                (\bm{\sigma}_\text{int} - \bm{\sigma}_\text{ext})
                \bm{\cdot \hat{n}} 
                \right]_{r = R}
                \bm{\hat{n}}
            $
            ,
            in the normal component to be given by:
            \begin{align}
                \Delta \bm{\Sigma}_\mu
                &=
                -6\,\mu \,
                \bm{\hat{n} \cdot E}^\infty \bm{\cdot \hat{n}}
                \
                \bm{\hat{n}}
                .
            \end{align}
            In the limit of infinite surface tension ($\sigma \to \infty$), the discontinuity, $\bm{\Sigma}_\mu$, only leads to an infinitesimally small change in the droplet's shape.
            However for large but finite surface tension ($\sigma \gg 1$), $\bm{\Sigma}_\mu$ will lead to a correspondingly small but finite perturbation to the droplet's shape. 
            Let us suppose that the deformed droplet have the nearly circular shape given by:
            \begin{align}
                r(\bm{x})
                &=
                R
                \left( 
                    1
                    +
                    \frac{2\epsilon}{\dot{\gamma}}\,
                    \bm{\hat{n} \cdot E}^\infty \bm{\cdot \hat{n}}
                \right),
            \end{align}
            where $\epsilon$ is a small dimensionless perturbation parameter.
            Noting that the capillary stress, $\bm{\Sigma}_\sigma$, at a point $\bm{x}$ on the interface is given by $\sigma \kappa \bm{\hat{n}}$, where $\sigma$ is the coefficient of surface tension and $\kappa$ is the mean curvature of the surface at that point, we find that for the nearly circular surface $\bm{\Sigma}_\sigma = -(\sigma/R)\bm{\hat{n}} + \Delta \bm{\Sigma}_\sigma$, where the perturbation, $\Delta \bm{\Sigma}_\sigma$, to the capillary stress is given by:
            \begin{align}
                \Delta \bm{\Sigma}_\sigma
                &=
                -\frac{6\,\epsilon}{\dot{\gamma}}
                \frac{\sigma}{R}
                \bm{\hat{n} \cdot E}^\infty \bm{\cdot \hat{n}}
                \
                \bm{\hat{n}}
                +
                \mathcal{O}(\epsilon^2).
            \end{align}
            The small deformation theory approximation is then made, in which the equilibrium shape of the droplet is assumed to be such that the discontinuity, $\Delta \bm{\Sigma}_\mu$, in the normal component of the viscous stress be equal and opposite to the change, $\Delta \bm{\Sigma}_\sigma$, in capillary stress.
            This leads us to identify the small parameter $\epsilon$ with the dimensionless quantity $\mu \dot{\gamma} R/\sigma$, which we recall as being the capillary number, Ca, of the system.
            The Taylor deformation parameter is defined as \cite{Taylor1932, Taylor1934}:
            \begin{align}
                D_T
                &=
                \frac{l-b}{l+b}
            \end{align}
            where $l$ is the longest dimension of the droplet and $b$ is its shortest.
            As in the 3D case, we expect the steady-state value, $D_T^\infty$, to satisfy:
            \begin{align}
                D_T^\infty
                &=
                g(\lambda)\,
                \text{Ca}
                \label{eqn:deformation_capillary}
            \end{align}
            where $g(\lambda)$ is a function to be determined.
            For our nearly circular droplet, we have:
            \begin{align}
                l
                &=
                R
                \left( 
                    1
                    +
                    \text{Ca}
                \right)
                \\
                b
                &=
                R
                \left( 
                    1
                    -
                    \text{Ca}
                \right).
            \end{align}
            Thus in the limit of large surface tension ($\text{Ca} \to 0$), we find:
            \begin{align}
                D_T^\infty
                &=
                \text{Ca}
                \label{eqn:Dt=Ca}
            \end{align}
            Hence, $g(\lambda) = 1$. 
            This result is coincides exactly with results from other works in the asymptotic limit $\text{Ca} \to 0$.
            In particular, it coincides with the work of Buckmaster and Flaherty~\cite{buckmaster1973bursting} for droplets ($\lambda > 0$) in purely extensional flow, as well as the work of Richardson~\cite{richardson1968two, richardson1973two} for bubbles ($\lambda = 0$) in simple shear.
            The work of Toose et. al.~\cite{toose1995boundary} provides numerical simulations of viscous cylinders in purely extensional flow, whose results are obey Eq.\eqref{eqn:Dt=Ca} asymptotically in the limit Ca $\to 0$.
            It will be noted that unlike in the 3D case, where $g^{3D}(\lambda) = (19\lambda + 16)/(16\lambda + 16)$, in the 2D case, the constant of proportionality is independent of the viscosity ratio, $\lambda$, between the inner and outer fluid.
            This is due to the fact that whilst in the 3D case, the stress jump depends on $\lambda$, in the present 2D case, it does not.
            We emphasize that the range $1 \leq g^{3D}(\lambda) \leq 19/16 \approx 1.19$ represents a deviation of nearly 20\% across different viscosity ratios. This variation is appreciable, particularly in the context of developing numerical simulations and the comparison of 2D and 3D theories and computations.
        
    \section{Numerical Simulations}
        \label{sec:numerics}
        We proceed by discussing our direct numerical simulations (DNS) of the full equations of motion and compare them with our analytical results.
        Our simulations were carried out using Basilisk C, an open-source language developed by S. Popinet and collaborators \cite{Popinet2003, Popinet2009, Popinet2015} for solving differential equations on adaptive Cartesian grids. 
        In particular, we use their Navier-Stokes and Volume of Fluids (VOF) solvers, which have been widely used and validated in other published works \cite[see][ for extended discussions on such comparisons]{Popinet2009, sanjay2022viscous}.
        Also, see Appendix \ref{appendix:numerics} for more information on how the simulations were implemented and validated.  
    
        \subsection{Non-Deformable Droplets}
    Firstly, we perform simulations in the capillarity-dominated regime $\text{Ca} \ll 1$ (more specifically $\text{Ca} = 0.01$), in which the deformation of the droplet is negligible.
    In this limit, the solution found in our simulations should be consistent with the analytical solution found in Section \ref{sec:analytics}.
    A more complete demonstration of the consistency between the numerical and analytical solutions can be found in Appendix \ref{appendix:numerics}.

    \begin{figure}[h]
        \centering
        \begin{subfigure}{0.49\textwidth}
            \includegraphics[width=\textwidth]{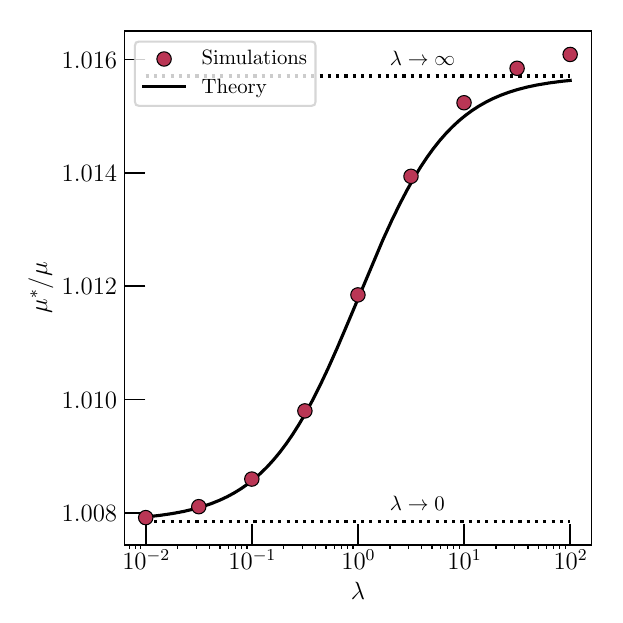}
            \caption{}
            \label{fig:einstein_viscosity}
        \end{subfigure}
        \hfill
        \begin{subfigure}{0.49\textwidth}
            \includegraphics[width=\textwidth]{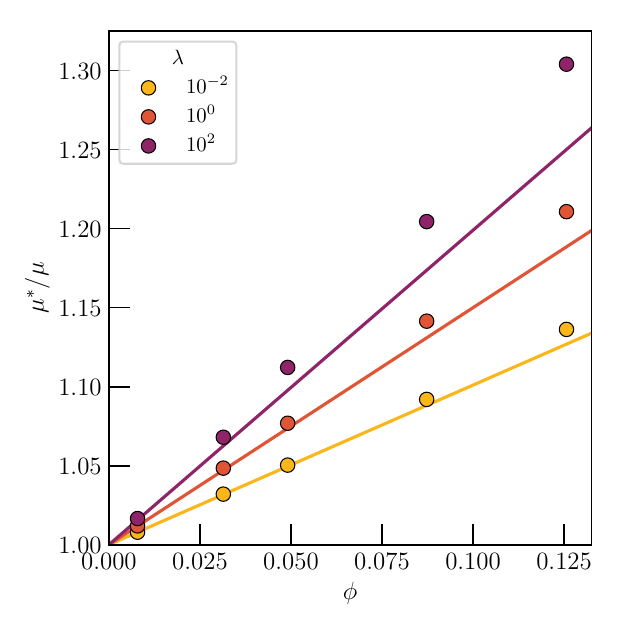}
            \caption{}
            \label{fig:einstein_viscosity_phi}
        \end{subfigure}
        \caption
        {
            (a) The apparent viscosity, $\mu^*$, of the system as a function of the viscosity ratio, $\lambda$.
            The black line shows the result derived from the analytical solution.
            The red circular points show the results obtained from numerical simulations.
            Here, $L = 20R$ such that the area fraction is $\phi \approx 0.008$.
            Also shown are the asymptotic limits $\lambda \to 0$ and $\lambda \to \infty$.
            (b) The apparent viscosity, $\mu^*$, of the system as a function of area fraction, $\phi$, of the suspended phase.
            The viscosity ratio, $\lambda$, is varied from $10^{-2}$ to $10^{2}$. The coloured circular points show data from the simulations, while the lines show the predictions from the theory.
            The agreement between simulations and theory improves in the limit $\phi \to 0$.
        }
    \end{figure} 
    
    In figure \ref{fig:einstein_viscosity}, we plot the apparent viscosity, $\mu^*$, of the droplet under the shear system for different values of the viscosity ratio, $\lambda$.
    The value of $\mu^*$ obtained from simulations via the formula in Eq. \eqref{eqn:einstein_dissipation} is compared with the analytical result, Eq. \eqref{eqn:einstein_viscosity}, with the volume (area) fraction used to evaluate the formula being $\phi = \pi R^2/L^2 \approx 0.008$.
    The numerical results are consistent with the analytical solution to within a good degree of accuracy over a large range of $\lambda$, namely for $0.01 \leq \lambda \leq 100$.
    In figure \ref{fig:einstein_viscosity_phi}, $\mu^*$ is plotted against $\phi$ while varying $\lambda$. 
    Here, $\phi$ was varied by altering dimensionless size, $L/R$, of the domain.
    The analytical result, Eq. \eqref{eqn:einstein_viscosity}, predicts a linear relationship between $\mu^*$ and $\phi$ in the dilute limit, $\phi \to 0$, in which interactions between the droplet, its neighbours and the boundary are negligible. 
    Eq. \eqref{eqn:einstein_viscosity} fits the numerical results well in this limit, but tends to underestimate $\mu^*$ for larger $\phi.$
    This is consistent with the behaviour of real emulsions where the value of $\mu^*$ generally exceeds the values predicted by linear theories (see, for example, \cite{Stickel}).
    It should be noted that, for the configuration used in this study, the area fraction, $\phi$, and the geometric confinement, $L/R$, are coupled; consequently, it is not possible to attribute the observed deviations solely to droplet–droplet interactions arising from periodicity, rather than confinement effects.
    The discrepancies at larger $\lambda$ ($\lambda > 10$) are believed to be numerical errors due to limitations in the maximum grid refinement.
    This is discussed in greater detail in Appendix \ref{appendix:numerics}.
    \begin{figure}[h]
        \centering
        \begin{subfigure}{0.28\textwidth}
            \includegraphics[width=\textwidth]{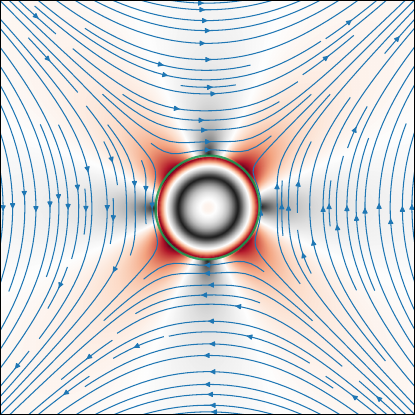}
            \caption{}
            \label{fig:first}
        \end{subfigure}
        \hfill
        \begin{subfigure}{0.28\textwidth}
            \includegraphics[width=\textwidth]{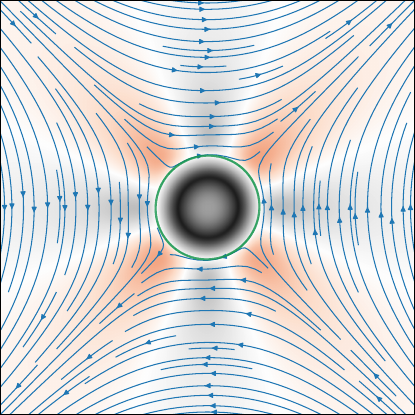}
            \caption{}
            \label{fig:second}
        \end{subfigure}
        \hfill
        \begin{subfigure}{0.28\textwidth}
            \includegraphics[width=\textwidth]{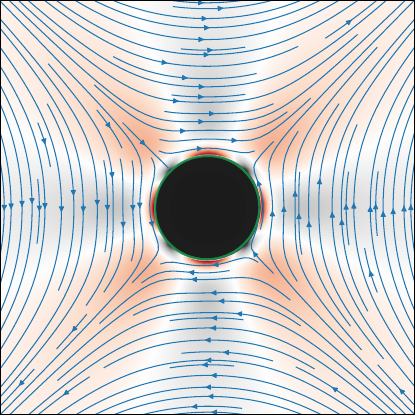}
            \caption{}
            \label{fig:third}
        \end{subfigure}
        \captionsetup[subfigure]{labelformat=empty}
        \begin{subfigure}{0.1\textwidth}
            \includegraphics[width=\textwidth]{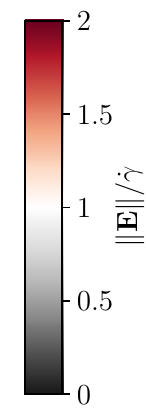}
            \caption{}
        \end{subfigure}
                
        \caption
        {
            The norm, $\norm{\bm{E}}$ of the strain rate tensor for non-deformable droplets ($\text{Ca} =  0.01$) with different values of $\lambda$.
            Panels $(a), (b)$ and $(c)$ show $\lambda = 0.01, 1$ and $100$ respectively.
            In each panel Re = 0.01 and $L = 20R$.
            The streamlines depict the vector field, $\bm{u} - \bm{R}^\infty \bm{\cdot x}$, that is the velocity field with the solid-body rotational part of the undisturbed flow subtracted.
        }
        \label{fig:non_deformable_droplets}
    \end{figure}

    In figure \ref{fig:non_deformable_droplets}, we plot the norm, $\norm{\bm{E}} = \sqrt{\bm{E : E}/2}$, of the strain rate tensor, $\bm{E} = (\bm{\nabla u} + (\bm{\nabla u})^T)/2$, for non-deformable droplets for different values of the viscosity ratio, $\lambda$.
    In each subfigure, streamlines are superimposed indicating the vector field $\bm{u} - \bm{R}^\infty \bm{\cdot x}$.
    This vector field exhibits a quadrupolar structure, which retains the four-fold rotational symmetry of the extensional part, $\bm{E}^\infty \bm{\cdot x}$, of the undisturbed flow.
    On the interior, we see that in the limit $\lambda \to \infty$, the strain rate tends to zero, which is consistent with the interpretation of that limit being the solid particle limit.

    \subsection{Deformable Droplets}
        \begin{figure}[h!]
            \centering
            \begin{subfigure}{0.28\textwidth}
                \includegraphics[width=\textwidth]{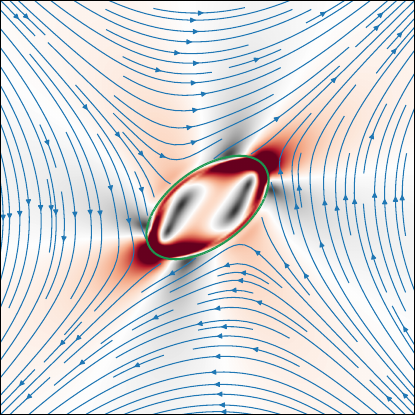}
                \caption{}
                \label{fig:first}
            \end{subfigure}
            \hfill
            \begin{subfigure}{0.28\textwidth}
                \includegraphics[width=\textwidth]{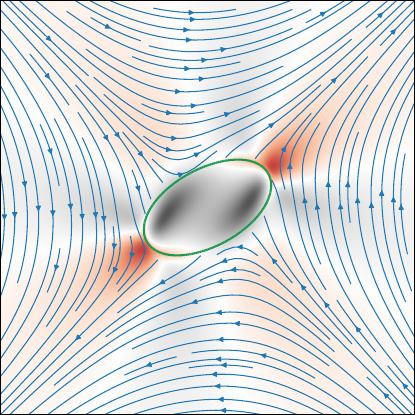}
                \caption{}
                \label{fig:second}
            \end{subfigure}
            \hfill
            \begin{subfigure}{0.28\textwidth}
                \includegraphics[width=\textwidth]{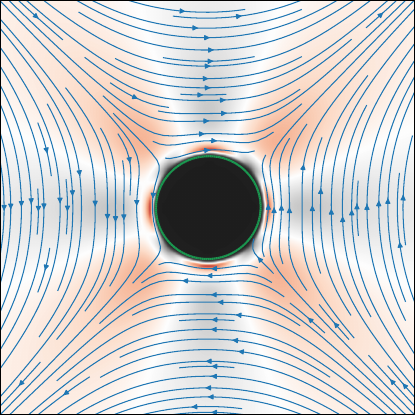}
                \caption{}
                \label{fig:third}
            \end{subfigure}
            \captionsetup[subfigure]{labelformat=empty}
            \begin{subfigure}{0.1\textwidth}
                \includegraphics[width=\textwidth]{figures/colourbar.pdf}
                \caption{}
            \end{subfigure}  
            \caption
            {
                The norm, $\norm{\bm{E}}$, of the strain rate tensor for deformable droplets ($\text{Ca} =  0.3$) with different values of $\lambda$.
                Panels $(a), (b)$ and $(c)$ show $\lambda = 0.01, 1$ and $100$, respectively.
                In each panel Re = 0.01 and $L = 20R$.
                The streamlines depict the vector field, $\bm{u} - \bm{R}^\infty \bm{\cdot x}$, that is the velocity field with the solid-body rotational part of the undisturbed flow subtracted.
            }
            \label{fig:deformable_droplets}
        \end{figure}
\begin{figure}[h!]
            \centering
            \begin{subfigure}{0.49\textwidth}
                \includegraphics[width=\textwidth]{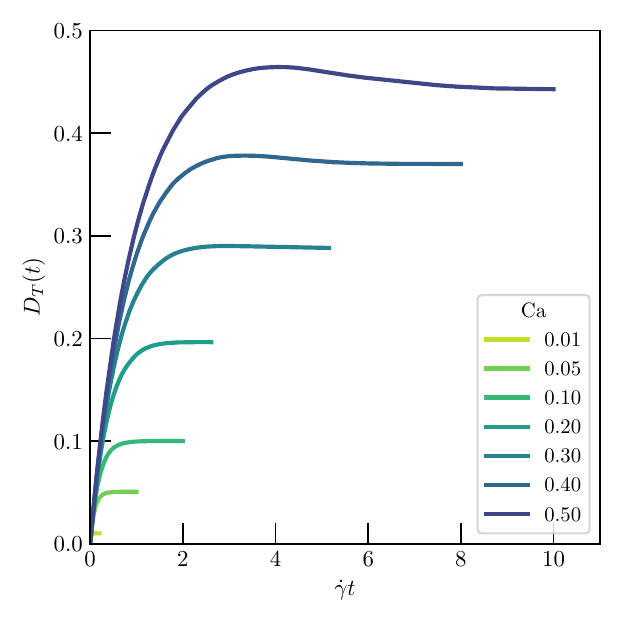}
                \caption{}
            \end{subfigure}  
            \hfill
            \begin{subfigure}{0.49\textwidth}
                \includegraphics[width=\textwidth]{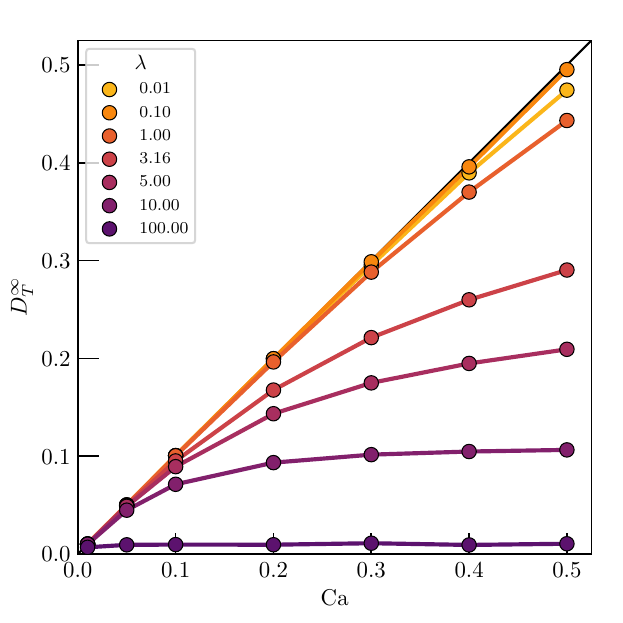}
                \caption{}
            \end{subfigure}
            \caption
                {
                    The dependence of the Taylor deformation parameter, $D_T$, on the viscosity ratio, $\lambda$ and the capillary number, Ca.
                    Panel $(a)$ shows $D_T$ as a function of time for $\lambda = 1.0$ and different Ca.
                    Panel $(b)$ shows the steady-state value, $D_T^\infty$, of $D_T$ for varying Ca and $\lambda$.
                    The small deformation theory predicts $D_T^\infty = \text{Ca}$ for small Ca, independently of the value of $\lambda$.
                    This result holds in the linear regime throughout the tested range $0.01 \leq \lambda \leq 100$. 
                    It will be noted that the linear regime ends for smaller Ca as $\lambda$ increases, consistent with convergence to solid-particle behaviour.
                }
                \label{fig:deformation_theory}
        \end{figure}

        We also perform simulations in the moderate capillarity regime, $0 < \text{Ca} < 1$, in which the droplet undergoes finite deformations without rupturing (see Figure~\ref{fig:deformable_droplets}).
        In each simulation, the initially circular droplet immediately begins to deform.
        The system is allowed to reach a steady state, and the corresponding value, $D_T^\infty$, of the Taylor parameter is then recorded.
        Based on the calculation in Section \ref{sec:analytics}, we expect that $D_T^\infty = \text{Ca}$, provided Ca is small enough.
        Figure \ref{fig:deformation_theory} shows a comparison between our numerical results and this prediction.
        The numerical data is consistent with the small deformation theory for a large range of $\lambda$ ($0.01 \leq \lambda \leq 100$).
        However, the extent in Ca of the linear regime decreases with increasing $\lambda$.
        In particular, for large $\lambda$, we observe that the value of $D_T$ saturates at a finite value in the limit $\text{Ca} \gg 1$.
        This behaviour is also encountered in the 3D case \cite{Rallison_1980, Cox1969, BarthsBiesel1973}.
        In Appendix \ref{appendix:large_lambda}, we show that this arises due to a qualitatively different steady state which is independent of Ca.
        In particular, it can be shown that for large $\lambda$, $D_T^\infty \approx 1/(\lambda + 1)$.
        The transition to this regime occurs at $\text{Ca} \approx 1/(\lambda + 1)$.
        For extremely large $\lambda$, in particular $\lambda = 100$, we observe that the droplet undergoes negligible deformation and undergoes a constant solid body rotation, which is consistent with our expectation that in the limit $\lambda \to \infty$, we should recover the properties of a rigid particle.

\pagebreak

    \section{Conclusions}
        \label{sec:conclusions}
        We have derived an analytical solution for the two-dimensional droplet under shear problem using a 2D Lamb approach.
        We have shown that the apparent viscosity of a dilute emulsion is
        $\mu^* = \mu (1 + f(\lambda) \phi) + \mathcal{O}(\phi^2)$
        where 
        $f(\lambda) = (2\lambda + 1)/(\lambda + 1)$
        ,
        $\lambda$
        is the ratio of the viscosity of the fluid in the droplet to the viscosity of the matrix
        and $\phi$ is the volume (area) fraction of the suspended phase.
        We have shown that to leading order, the steady-state Taylor parameter is given by
        $D_T^\infty = g(\lambda) \text{Ca} + \mathcal{O}(\text{Ca}^2)$ 
        where $g(\lambda) = 1$ independently of the viscosity ratio.
        This region of validity of this equation becomes smaller for larger $\lambda$.
        The results establish a theoretical foundation, validated by the numerical simulations, serving as a benchmark for computational models. 
        We have not discussed larger deformations, time-dependence, or the deformation of droplets with larger $\lambda$; these could be immediate extensions of the work.
        Other future work could extend the analysis to include droplet-droplet interactions, interactions with the boundaries, the effects of alternative boundary conditions, as well as non-dilute systems containing many droplets where collective effects become important.
        Incorporating additional rheological features, such as elasticity~\cite{Guido2011-ug, AGGARWAL_SARKAR_2007, AGGARWAL_SARKAR_2008, AGGARWAL200819}, plasticity~\cite{Li2000_BI}, elasto-viscoplasticity~\cite{Izbassarov,francca2024elasto}, surfactants and surface rheology~\cite{stone1990effects,li1997effect,narsimhan2019shape}, and odd viscosity~\cite{Avron1998, Fruchart, francca2025odd} could further enrich the model.
          \clearpage 
            
\section*{Appendices}
    \appendix

    \section{The Lamb Solution for Two-Dimensional Stokes Flow}
        \label{appendix:analytics}
        We will derive a general solution to the 2D Stokes equations. 
        \begin{align}
            \bm{\nabla \cdot u}
            &=
            0
            \label{eqn:appendix_cont}
            \\
            \bm{\nabla} p
            &=
            \mu \nabla^2 \bm{u}.
            \label{eqn:appendix_stokes}
        \end{align}
        It can be shown that for any flow field $\bm{u}$ satisfying both the Stokes equation \eqref{eqn:appendix_stokes} and the continuity equation \eqref{eqn:appendix_cont}, the pressure function, $p$, corresponding to it must be a harmonic function.
        It then follows that $p$ can be written as a sum of solid circular harmonics:
        \begin{align}
            p 
            &=
            \sum_{n=-\infty}^\infty
            p_n,
        \end{align}
        where the subscript, $n$, indicates that the solid circular harmonic, $p_n$, is of degree $n$ in the radial coordinate, $r$.
        In view of the Stokes equations, it follows that the flow field, $\bm{u}$, should also be a sum of terms with different degrees in $r$:
        \begin{align}
            \bm{u}
            &=
            \sum_{n=-\infty}^\infty
            \bm{u}_n.
        \end{align}
        We shall adopt the convention that the function $\bm{u}_n$ denote a function of degree $n+1$ in $r$, as it then follows that for any pair 
        $\{\bm{u}_n, p_n\}$,
        the corresponding Stokes equation for that pair must be satisfied:
        \begin{align}
            \bm{\nabla} p_n &= \mu \nabla^2 \bm{u}_n.
        \end{align}
        The task of solving the Stokes equation has hereby been reduced to the following problem: given an arbitrary solid circular harmonic, $p_n$, of degree $n$, what form must $\bm{u}_n$ take, such that the pair, $\{\bm{u}_n, p_n\}$, satisfy the Stokes equations. 
        To this end, we suggest the following ansatz:
        \begin{align}
            \bm{u}_n
            &=
            \bm{u}'_n
            +
            A_n r^2 \bm{\nabla} p_n
            +
            B_n r^{2n + 2} 
            \bm{\nabla} \left(\frac{p_n}{r^{2n}}\right)
        \end{align}
        for some particular choice of $A_n$ and $B_n$, and where $\bm{u}'_n$ is an arbitrary harmonic function of degree $n + 1$.
        This can also be written in the following form:
        \begin{align}
            \bm{u}_n
            &=
            \bm{u}'_n
            +
            (A_n + B_n)r^2 \bm{\nabla} p_n
            -
            2n B_n p_n \bm{x}
            \label{eqn:appendix_ansatz}
            .
        \end{align}
        Substituting the ansatz \eqref{eqn:appendix_ansatz} first into the continuity equation \eqref{eqn:appendix_cont} and then into the momentum equation \eqref{eqn:appendix_stokes}, it can be shown that the correct expressions for $A_n$ and $B_n$ are:
        \begin{align}
            A_n
            &=
            \frac{1}{4\mu n}
            \\
            B_n
            &=
            \frac{1}{4\mu n (n+1)}
        \end{align}
        It can also be shown that any harmonic function, $\bm{u}_n'$, of degree $n+1$ satisfying the continuity equation can be expressed in the form: 
        \begin{align}
            \bm{u}'_n
            &=
            \bm{\nabla} \phi_{n+1} 
            +
            \bm{\nabla} \times (\bm{x}\chi_n),
        \end{align}
        where $\phi_{n}$ and $\chi_n$ are arbitrary solid circular harmonics of degree $n$.
        By linearity, we can superpose the functions, $\bm{u}_n$, to obtain a general solution to the two-dimensional Stokes equations.
        \begin{align}
            \bm{u}
            &=
            \sum_{n=-\infty}^{\infty}
            \left\{
            \bm{\nabla} \phi_n
            + 
            \bm{\nabla} \times (\bm{x}\chi_n)
            +
            \frac{n+2}{4 \mu n(n+1)}
            r^2 \bm{\nabla} p_n
            -
            \frac{1}{2\mu(n+1)}p_n \bm{x}
            \right\}
            \\
            p 
            &=
            \sum_{n=-\infty}^\infty
            p_n
        \end{align}
        By a simple computation, we obtain the vorticity of this flow:
        \begin{align}
            \bm{\omega}
            &=
            \sum_{n= - \infty}^{\infty}
            \left\{
            \frac{1}{\mu n}
            (\bm{x} \times \bm{\nabla}) p_n
            +
            n \bm{\nabla} \chi_n
            \right\}.
        \end{align}
        Now that we have a general solution to the Stokes equations, solving these equations for any given system is simply a matter of selecting the appropriate forms of the solid circular harmonics, $p_n$, $\phi_n$ and $\chi_n$, compatible with the boundary conditions imposed at the surface of the droplet.
        These boundary conditions will involve the flux of velocity, vorticity and stress through the surface.
        We therefore state the following general results:
        \begin{align}
            \bm{x \cdot u}
            &=
            \sum_{n=-\infty}^{\infty}
            \left\{
            n \phi_n
            +
            \frac{n}{4 \mu(n+1)}
            r^2 p_n
            \right\}
            \\
            \bm{x \cdot \omega}
            &=
            \sum_{n= - \infty}^{\infty}
            \left\{
            n^2 \chi_n
            \right\}
            \\
            \bm{x \cdot \sigma}
            &=
            \sum_{n= - \infty}^{\infty}
            \bigg\{
            2\mu(n-1)\bm{\nabla} \phi_n 
            +
            \mu(n-1) \bm{\nabla} \times (\bm{x}\chi_n)
            +
            \frac{1}{2}r^2
            \bm{\nabla} p_n
            -
            p_n\bm{x}
            \bigg\}
            \\
            \bm{x \cdot \sigma \cdot x}
            &=
            \sum_{n= - \infty}^{\infty}
            \bigg\{
            2\mu n(n-1)\phi_n 
            +
            \frac{1}{2}
            (n - 2)
            r^2 p_n
            \bigg\}.
        \end{align}

    \clearpage
    
    \section{Highly-Viscous Droplets $(\lambda \gg 1)$}
        \label{appendix:large_lambda}
        In Section \ref{sec:analytics}, it was shown that for small Ca,
        $D_T^\infty = g(\lambda) \text{Ca} + \mathcal{O}(\text{Ca}^2)$.
        In Section \ref{sec:numerics}, this prediction was validated against numerical simulations, which revealed a linear regime whose extent decreases with increasing $\lambda$.
        Here, we estimate the extent of this regime and show that it holds for $\text{Ca} < 1/(\lambda + 1)$.
        
        In the final step of the derivation in Section \ref{sec:analytics}, the constants $a_{-2}$, $b_{-2}$, $a_2$ and $b_2$ were determined by imposing boundary conditions at the interface.
        As is standard at a fluid-fluid interface, we imposed the continuity of the velocity field and the tangential component of the stress tensor.
        These boundary conditions led to three independent equations relating $a_{-2}, b_{-2}, a_{2}$ and $b_{2}$.
        The system of equations was then closed by imposing the no-penetration condition.
        This choice of boundary conditions leads necessarily to a discontinuity in the normal component of the stress tensor at the interface.
        However in the limit, $\text{Ca}\to 0$, the jump in normal stress across the interface can be balanced by an infinitesimal perturbation to the droplet's shape, which in turn leads to an infinitesimal perturbation to the flow.
        Thus the solution could be considered exact in the limit, $\text{Ca}\to 0$.
        
        If instead we force continuity of the normal component of the stress, the no-penetration condition cannot be satisfied, resulting in a non-zero normal component of the velocity at the interface.
        We now show the normal component scales as $1/\lambda$, which for large $\lambda$ can be balanced by a small shape perturbation.
        This in turn leads to a different small deformation theory, independent of Ca, which we use to estimate the extent of the linear regime.
        
        \subsection{Approximate Analytical Solution for $\lambda \gg 1$}
            Imposing continuity of the velocity and of both tangential and normal stresses at the interface yields
            \begin{align}
                a_2 
                &=
                0\\
                b_2 
                &=
                -\frac{\lambda - 1}{2(\lambda + 1)}\\
                a_{-2} 
                &= 
                -\frac{4(\lambda - 1)}{\lambda + 1}\\
                b_{-2} 
                &= 
                - \frac{\lambda - 1}{2(\lambda +1)}
            \end{align}
            The corresponding flow and pressure fields are
            \begin{align}
                \bm{u}^{\text{int}}
                -
                \bm{u}^\infty
                &=
                -\frac{\lambda - 1}{\lambda + 1} \bm{E}^\infty \bm{ \cdot x}
                \label{eqn:u_int_viscous}
                \\
                p^{\text{int}} - p^\infty
                &=
                0
                \label{eqn:p_int_viscous}
                \\
                \bm{u}^{\text{ext}}
                -
                \bm{u}^\infty
                &=
                -\frac{2(\lambda - 1)}{\lambda + 1}
                \left( 
                \frac{R}{r}
                \right)^2
                \frac{\bm{x \cdot}  \bm{E}^\infty \bm{\cdot x}}{r^2}\bm{x}
                -
                \frac{\lambda - 1}{\lambda + 1}
                \left( 
                \frac{R}{r}
                \right)^4
                \left( 
                \bm{E}^\infty \bm{ \cdot x}
                -
                2 \frac{\bm{x \cdot}  \bm{E}^\infty \bm{\cdot x}}{r^2} \bm{x}
                \right)
                \label{eqn:u_ext_viscous}
                \\
                p^{\text{ext}} - p^\infty
                &=
                -\frac{4(\lambda - 1)\mu }{\lambda + 1} 
                \left( \frac{R}{r} \right)^2
                \frac{\bm{x \cdot}  \bm{E}^\infty \bm{\cdot x}}{r^2}
                \label{eqn:p_ext_viscous}
            \end{align}
            As before, in the limit $\lambda \to \infty$, this again reduces to the classical solution for extensional flow around a rigid cylinder \cite{BRADY1983}.
        
        \subsection{Small Deformation Theory for $\lambda \gg 1$}
            To derive the small-deformation theory, we use 2D polar coordinates and write $\bm{u} = u_r(r,\theta) \, \bm{\hat{r}} + u_\theta(r, \theta) \, \bm{\hat{\theta}}$ where:
            \begin{align}
                u_r (r,\theta)
                &= 
                -\frac{1}{\lambda + 1}
                \dot{\gamma} \, r 
                \sin(2\theta)
                \\
                u_\theta (r,\theta)
                &= 
                -\frac{1}{\lambda + 1}
                \dot{\gamma} \, r 
                \cos(2\theta)
                -
                \frac{\dot{\gamma}\,r}{2}
            \end{align}
            We then propose the following form for the nearly circular droplet:
            \begin{align}
                r(\theta)
                &=
                R(1 + \epsilon \, \xi(\theta))
            \end{align}
            where $\xi(\theta)$ is a dimensionless perturbation to the shape.
            Since we assume $\lambda$ to be large, we take $\epsilon$ to be of the order $1/\lambda$.
            Its precise form is determined by the boundary conditions at the interface.
            We impose the kinematic condition that the interface is advected by the flow, which reads
            $DF(r, \theta)/Dt = 0$ 
            where 
            $D/Dt = \partial/\partial t + \bm{u \cdot \nabla}$ is the material derivative and $F(r, \theta) = r - R(1 - \epsilon \, \xi(\theta))$
            defines the interface via $F(r,\theta)=0$.
            In steady state, $\partial/\partial t = 0$ and the condition reduces to $\bm{u}\cdot\bm{\nabla}F = 0$.
            Evaluating this at the perturbed interface and retaining only terms of order $1/\lambda$ yields
            $\epsilon = 1/(\lambda + 1)$ and $\xi(\theta) = \cos(2\theta)$.
            It follows that the steady-state Taylor parameter, $D_T^\infty$ is:
            \begin{align}
                D_T^\infty 
                &= 
                \frac{1}{\lambda + 1}
                \label{eqn:deformation_viscous}
            \end{align}
            Since there is no dependence on Ca, it follows that droplets still reach a steady state even in the weak capillarity regime ($\text{Ca} \gg 1$).
            However, their orientation differs from the strong-capillarity case: instead of having the major axis at $45^\circ$ anti-clockwise from the horizontal, a highly viscous droplet aligns its long axis with the flow direction.
            This behaviour is also encountered in the 3D case \cite{Taylor1934}).
            Comparing Eqs. \eqref{eqn:deformation_capillary} and \eqref{eqn:deformation_viscous}, we estimate the transition between the capillarity-dominated regime and the highly viscous regime to occur at $\text{Ca} \sim 1/(\lambda + 1)$.
            \begin{figure}[h!]
                \centering
                \begin{subfigure}{0.49\textwidth}
                    \includegraphics[width=\textwidth]{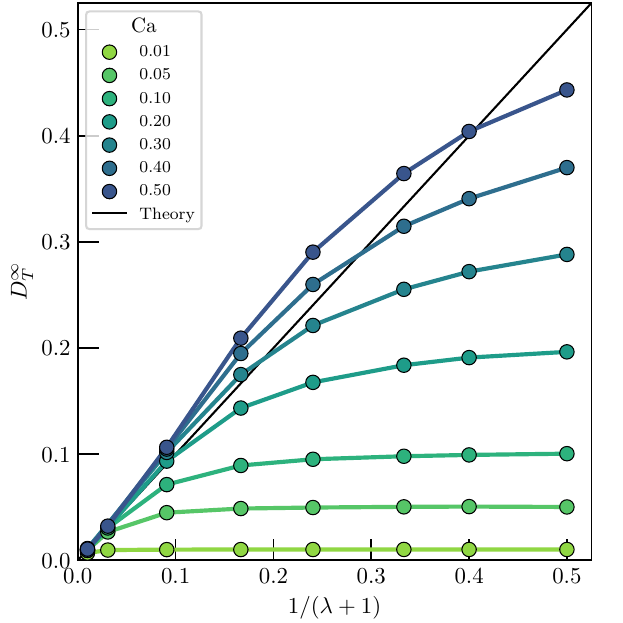}
                \end{subfigure}
                
                \caption
                    {
                        The dependence of the Taylor deformation parameter $D_T$ on the viscosity ratio $\lambda$ and the capillary number $\text{Ca}$.
                        The steady-state value $D_T^\infty$ is plotted against $1/(\lambda + 1)$ for different $\text{Ca}$.
                        The coloured circular symbols show the results obtained from numerical simulations.
                        The large $\lambda$ linear theory $D_T^\infty = 1/(\lambda + 1)$ is shown as a solid black line.
                        When $1/(\lambda + 1) \ll \text{Ca}$, the large $\lambda$ linear theory applies.
                        When $1/(\lambda + 1) \gg \text{Ca}$, the capillarity-dominated regime $D_T^\infty = \text{Ca}$ is recovered.     
                    }
                
                    \label{fig:deformation_theory_high_viscous}
            \end{figure}

    \clearpage
    \section{The Dependence of Apparent Viscosity on $\text{Ca}$}
        \label{appendix:apaprent_Ca}
\begin{figure}[h]
        \centering
        \includegraphics[width=0.5\textwidth]{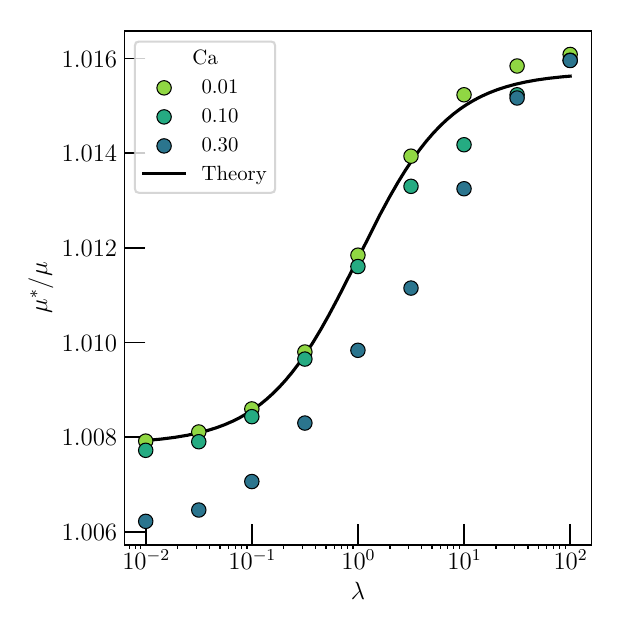}
        \caption{
            The apparent viscosity, $\mu^*$, of the droplet system under shear as a function of the viscosity ratio, $\lambda$, for different values of the capillary number, $\text{Ca}$.
            The coloured circular symbols show the results obtained from numerical simulations.
            The solid black line shows the theoretical prediction \eqref{eqn:apparent_2D}.
            Here, $L/R = 20$, corresponding to an area fraction $\phi \approx 0.008$.
        }
        \label{fig:einstein_viscosity_many_Ca}
    \end{figure}

        In section \ref{sec:analytics}, we derived the expression \eqref{eqn:apparent_2D} for the apparent viscosity, $\mu^*$, of a dilute two-dimensional emulsion of droplets.
        This derivation assumes the limit $\text{Ca} \to 0$.
        In this section, we examine the breakdown of this result at finite $\text{Ca}$.
        Figure \ref{fig:einstein_viscosity_many_Ca} shows the values of $\mu^*$ as a function of $\lambda$ for several values of $\text{Ca}$, as obtained from simulations, together with the theoretical prediction \eqref{eqn:apparent_2D}.
        For a given $\lambda$, the results show that $\mu^*$ decreases monotonically with increasing $\text{Ca}$.
        At $\text{Ca} = 0.01$ and $0.1$, the discrepancy with the theory is small, but already at $\text{Ca} = 0.3$ the deviation is of order $\phi$. 
 
    \clearpage
    \section{Validation of Numerical Simulations}
        \label{appendix:numerics}
            In this section we demonstrate the validity of our numerical results.
            In particular, we show that our results hold independently of the maximum level of mesh refinement, beyond refinement thresholds which we clearly define. Readers are referred to the original works of Popinet~\cite{Popinet2003,Popinet2009,Popinet2015} for details on the octree adaptive mesh refinement and the VOF method.
            The maximum level of mesh refinement, LEVEL, is defined such that the size, $L$, of the domain is $2^\text{LEVEL}$ times the size of the smallest cell.
            This implies that the maximum number of cells along the initial diameter of the droplet is $2^\text{LEVEL}\cdot 2R/L$, where $R$ is the initial radius of the droplet.
            For effectively non-deformable droplets (Ca = 0.01), where we are primarily concerned with the flow field, $\bm{u}$, and apparent viscosity, $\mu^*$, of the system, we fix LEVEL such that the number of cells over the diameter of the droplet is at least 200.
            This corresponds to $\text{LEVEL} = 11$ when $L/R = 20$.
            Meanwhile, for deformable droplets (Ca $>$ 0.01), where we are only concerned with the steady-state deformation parameter, $D_T^\infty$, we choose LEVEL such that the number of cells over the diameter of the droplet is at least approximately 50.
            This corresponds to $\text{LEVEL} = 9$ when $L/R = 20$.
            In the following, we demonstrate the sufficiency of these criteria.
            
\begin{figure}[h]
                \centering\includegraphics[width=\linewidth]{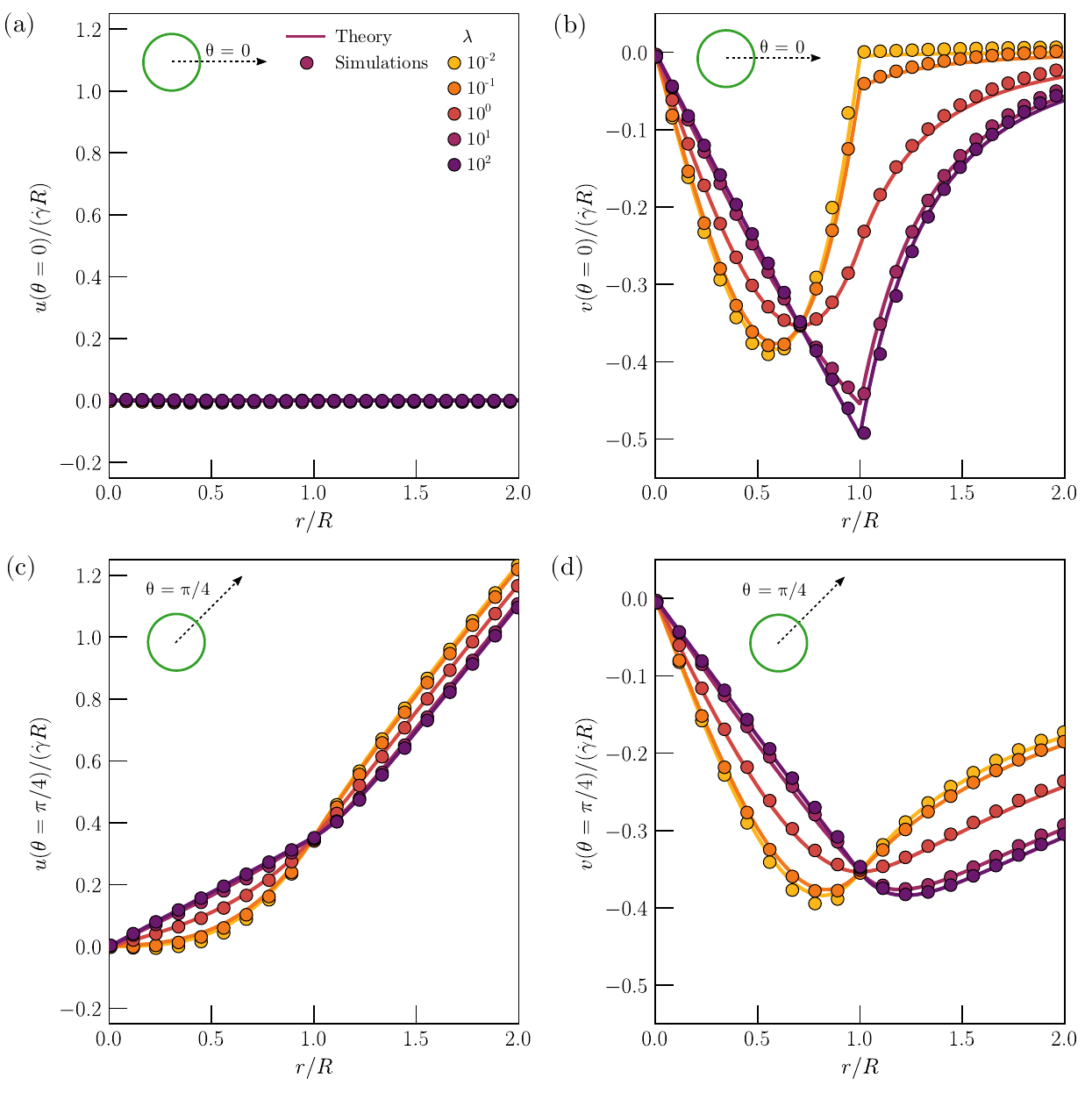}
                \caption{Validation of numerical simulations (circles) against the analytical solution Eqs. \eqref{eqn:u_int} and \eqref{eqn:u_ext} (lines). 
                Here, we fix $\text{Re} = 0.01$ to remain in the Stokes regime, and $\text{Ca} = 0.01$ such that the surface tension dominates and the droplet remains circular, consistent with the analytical solution.
                Here $L/R = 20$ and LEVEL = 11.
                The velocity components $u$ and $v$ are plotted along lines emanating from the centre of the droplet straight outwards at angles $\theta = 0$ and $\pi/4$ to the $x$-axis. 
                Our results show strong agreement for viscosity ratios in the range $10^{-2} \leq \lambda \leq 10^2$.}
                \label{fig:validate_velocity}
            \end{figure}

\begin{figure}[h]
                \begin{subfigure}{0.45\textwidth}
                    \includegraphics[width=\textwidth]{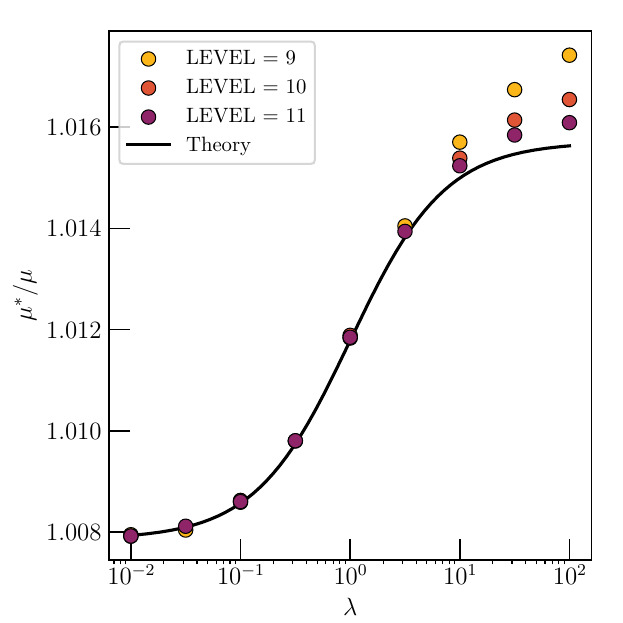}
                    \caption{}
                    \label{fig:validate_apparent}
                \end{subfigure}
                \begin{subfigure}{0.45\textwidth}
                    \includegraphics[width=\textwidth]{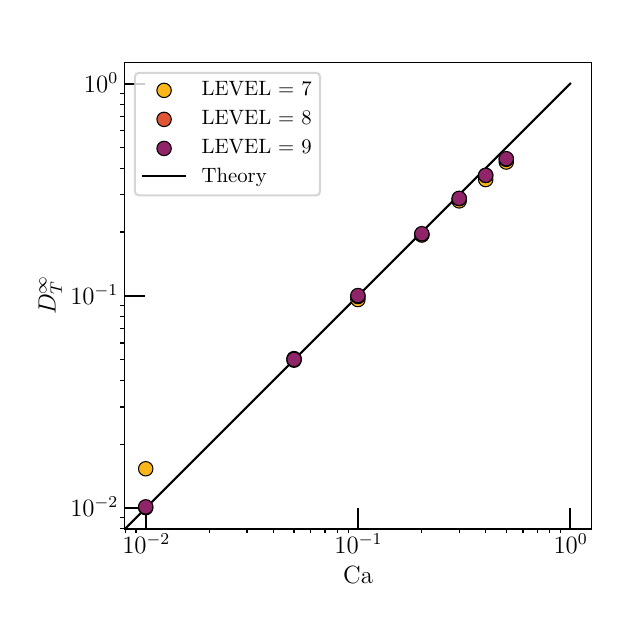}
                    \caption{}
                    \label{fig:validate_DT}
                \end{subfigure}
                \caption{
                    (a) Convergence of the apparent viscosity $\mu^*$ to the predicted value as the maximum grid refinement, LEVEL, increases. 
                    Here, we have fixed $\text{Re} = 0.01$, $\text{Ca} = 0.01$ and $L/R = 20$.
                    Already when $\text{LEVEL} = 9$ our results show good agreement with the theoretical values for $\lambda < 1$ however for $\lambda > 1$ discrepancies arise. 
                    These discrepancies however decrease as we increase the refinement to $\text{LEVEL} = 11$.
                    The computational cost of $\text{LEVEL} > 11$ was prohibitive.
                    (b)
                    Convergence of the steady-state Taylor deformation parameter, $D_T^\infty$, to the predicted value as the maximum grid refinement, LEVEL, increases. 
                    Here, we have fixed $\text{Re} = 0.01$ and kept $\text{Ca} \leq 0.5$.
                    We keep $L/R = 20$.
                    When $\text{LEVEL} = 9$, the measured values have converged, with $\text{LEVEL} = 7$ case already being consistent with the $\text{LEVEL} = 9$ case as well as the theoretical prediction, 
                    except for the Ca = 0.01 case, where a $\text{LEVEL} = 8$ is required to properly resolve the droplet deformation.
                    Thus our choice $\text{LEVEL} = 9$ guarantees a good resolution of $D_T^\infty$ over the full range $0.01 \leq \text{Ca} \leq 0.5$ used in this work.
                }
            \end{figure}

            In Figure \ref{fig:validate_velocity}, we demonstrate that in the case of an effectively non-deformable droplet (Re = 0.01, Ca = 0.01) with LEVEL = 11 and L/R = 20, the $x$- and $y$-components of the velocity field closely match the analytical solutions in Eqs. \eqref{eqn:u_int} and \eqref{eqn:u_ext}.
            In particular, we compare radial profiles of the $u$ and $v$ along the principle axes of the extensional flow, that is to say, along lines emanating from the centre of the droplet extending straight outwards at angles of $\theta = 0$ and $\pi/4$ from the horizontal $x$-axis.
            Our results show strong agreement for viscosity ratios in the range $10^{-2} \leq \lambda \leq 10^2$.
            In addition, we show in Figure \ref{fig:validate_apparent} that the numerically obtained values for the apparent viscosity, $\mu^*$, are consistent with the theoretical prediction in Eq. \eqref{eqn:apparent_2D}. 
            In particular, already when $\text{LEVEL} = 9$ our results show good agreement with the theoretical values for $10^{-2} \leq \lambda \leq 10^0$.
            However, for $\lambda > 1$ discrepancies arise. 
            These discrepancies, however, decrease as we increase the refinement from $\text{LEVEL} = 9$ to $\text{LEVEL} = 11$.
            It is expected that by increasing the level of refinement, the results would converge to the theoretical value, however the computational cost of simulations with $\text{LEVEL} > 11$ is currently prohibitively large.
            
            As for the values of $D_T^\infty$, in Figure $\ref{fig:validate_DT}$, we fix $\lambda = 1$ and $L/R = 20$ and show that already for LEVEL = 7, our numerical results are sufficiently converged and in close agreement with the theoretically predicted values, except in the case where Ca = 0.01, where the value of $D_T^\infty$ is too small to be measured accurately (also of the order 0.01).
            However, increasing the refinement to LEVEL = 9, we are able to accurately determine $D_T^\infty$ over the full range $0.01 \leq \text{Ca} \leq 0.5$ used in this work.
            
    \clearpage
    \section{Code Availability}
        All the code used in this study, along with the data produced by it, are openly available on GitHub at \cite{DATA}.
        \clearpage
        \printbibliography

\end{document}